\documentclass[%
 preprint,
 amsmath,amssymb,
 aps,
longbibliography
]{revtex4-1}

\usepackage{graphicx}
\usepackage{dcolumn}
\usepackage{bm}

\usepackage{adjustbox}
\usepackage{booktabs}
\usepackage{lipsum}
\usepackage{caption}
\usepackage{subcaption}
\usepackage{color}
\usepackage{calligra}
\usepackage[T1]{fontenc}
\newcommand\av{\color{red}}
\usepackage{upgreek}

\begin{document}


\title{Anomalous Diffusion in a Bench-Scale Pulsed Fluidized Bed}

\author{Jonathan E. Higham}
 \affiliation{University of Liverpool, School of Environmental Sciences, Department of Geography and Planning, Roxby Building, Liverpool, L69 7ZT, United Kingdom }
\author{Mehrdad Shahnam}
\author{Avinash Vaidheeswaran}%
\email{avinash.vaidheeswaran@netl.doe.gov}
\affiliation{%
 National Energy Technology Laboratory\\
 3610 Collins Ferry Rd\\
 Morgantown, West Virginia 26505, United States of America
 }%

\date{\today}

\begin{abstract}
We present our analysis on micro-rheology of a bench-scale pulsed fluidized bed, which represents a weakly confined system. Non-linear gas-particle and particle-particle interactions resulting from pulsed flow are associated with harmonic and sub-harmonic modes. While periodic structured bubble patterns are observed at the meso-scale, particle-scale measurements reveal anomolous diffusion in the driven granular medium. We use single-particle tracks to analyze ergodicity and ageing properties at two pulsing frequencies having remarkably different meso-scale features. The scaling of ensemble-averaged mean squared displacement is not unique. The distribution of time-averaged mean squared displacements is non-Gaussian, asymmetric and has a finite trivial contribution from particles in crowded quasi-static surroundings. Results indicate weak ergodicity breaking which along with ageing characterize the non-stationary and out-of-equilibrium dynamics.


\end{abstract}

\pacs{Valid PACS appear here}
\maketitle


\section{\label{sec:level1}Introduction}
Multiphase flows contain a broad range of spatio-temporal scales corresponding to complex non-linear dynamics \cite{Zhao2003,Blomgren2007,Fullmer2017,AV2017CES,AV2019,AV2020}. Fluidization is a notable example in which particles are suspended by an incoming stream of fluid, whereby they exhibit fluid-like behavior. Traveling kinematic waves manifest as bubbles which create spatial inhomogeneities in solids concentration. In particular, pulsed fluidized beds (PFBs) are characterized by recurring bubble patterns resulting from dynamical structuring and suppression of chaos compared to fluidized beds having non-perturbed inflow. PFBs have shown improved hydrodynamics by reducing or eliminating channeling or clumping of particles, and enhanced heat and mass transfer properties while being non-intrusive. Prior studies on PFBs are mostly restricted to meso- and macro-scale observations \cite{Pence98,Coppens2003,Coppens2012,Wu2017,Coppens2018,Higham2020}. Pulsing excites several interacting modes, and particle-level description is pivotal in elucidating some of the observed features. Trajectories of fluidized particles evolve depending on multiple factors such as external forcing, momentum exchange with carrier-phase, interactions with neighbors, material properties, and confinement, and transition through different states. For instance, particles in the vicinity of bubble wakes experience a greater acceleration compared to those near the distributor or other quasi-static regions.

Previous studies \cite{vanNoije98,barrat2002velocity,BenNaim05,Moka05,AV2017} have examined velocity fluctuations and reported deviations from ideal Brownian motion. The simplistic assumption of Maxwellian distribution breaks down quite easily in multi-particle systems, and results in anomalous diffusion where the ensemble-averaged mean squared displacement (MSD) is described by,

\begin{equation}
    \langle \mathrm{x}^{2}(\Delta) \rangle \sim \Delta^{\gamma}
\end{equation}

\noindent The process is sub-diffusive for $\gamma$ < 1 and super-diffusive for $\gamma$ > 1, both of which are observed in nature and engineering applications \cite{Solomon1993,Wong2004,Xie2008,Mattsson2009,Bronstein2009,Krapf2011,Kneller2011,Palombo2013}, while $\gamma$ = 1 describes Brownian diffusion. There also exist diffusive environments which cannot be described by a unique value of $\gamma$ and involve transition of regimes discussed above. Different sources of anomalous diffusion have been studied in the past which include continuous-time random walk (CTRW), fractional Brownian motion (FBM) and the motion governed by fractional Langevin equation (FLE), scaled Brownian motion (SBM), transport on a fractal support, and heterogeneous diffusion process (HDP). Previous analyses also include combining these parent processes such as CTRW-FLE \cite{Metzler2014-PRX} and SBM-HDP \cite{Metzler2015}, the latter was termed generalized diffusion process (GDP), where diffusivity follows,

\begin{equation}
\label{eq:GDP}
    \mathrm{D}(\mathrm{x},t) \sim (1+\beta) \mathrm{D}_{0} |\mathrm{x}|^{\alpha} t^{\beta}
\end{equation}

\noindent The above equation combines spatial and temporal dependence from the underlying HDP and SBM respectively. GDP is sub-diffusive ($\gamma$ < 1) when $\alpha > 2 \beta + 4$. 

Based on the physics of PFB \cite{Wu2017,Coppens2018,Higham2020,francia2020dynamically}, we hypothesize the system hosts a combination of parent processes discussed above, as will be shown in the remainder of this article. We also study the effect of ageing, i.e., time lapse after initializing experiments. It must be noted that PFB does not represent confinement in a strict sense. Boundaries or walls are present in the lateral directions which reflect particles after inelastic collisions, while the streamwise transport is constrained by balance between drag exerted by the carrier-phase and gravity. Hence, we describe the PFB as a weakly confined system. In addition, our unit is quasi-two-dimensional (quasi-2D), since the depth-wise extent is comparable to the size of bubbles, further verified by high-speed videos which reveal their span.

\section{\label{sec:level2}Experiments}
\begin{figure}[!htpb]
\begin{center}
\includegraphics[width=0.9\linewidth]{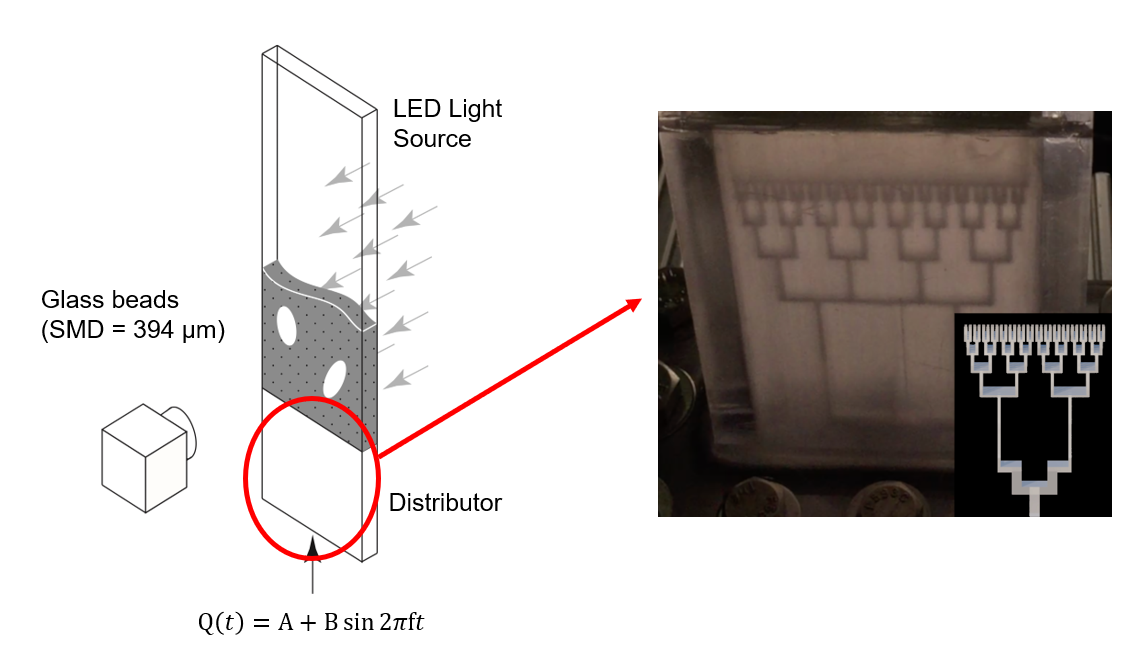}
\end{center}
\caption{Schematic of the PFB set-up used in this study. The zoomed-in image and inset show the frontal view and a rough sketch of fractal distributor (not drawn to scale).}
\label{fig:Schematic}
\end{figure}

The setup used for experiments (Figure ~\ref{fig:Schematic}) consists of a bench-scale test section with a cross-sectional area of 50$mm$ $\times$ 5$mm$. The unit was filled with 18$g$ of glass particles having a sauter mean diameter of 394$\mu m$ and a density of 2.5$g/cm^{3}$, classified under Geldart Group B \cite{Geldart1973}. The resulting static height was 50$mm$. Flow rate at the inlet was pulsed in the form of a sine wave,

\begin{equation}
    \mathrm{Q}(t) = \mathrm{A + B\: sin} (2 \pi \mathrm{f} t)
\end{equation}

\noindent where, the base flow rate, A = 2.6$l/min$. The corresponding velocity is higher than the minimum fluidization velocity, $U_{mf}$ which denotes the minimum velocity required to support the weight of solids. Details regarding the measurement of $U_{mf}$ can be found in \citet{AV-TRS2020}. The amplitude, B is set to 2.1$l/min$, and two pulsing frequencies are used, f=4$Hz$ and 6$Hz$. A fractal distributor was 3-D printed using a high-precision ultraviolet curing printer. High-speed videos were recorded at 300$Hz$ over a duration of 20$s$ using a 120$mm$ Nikon lens and Fastex IL5L sensor, and the unit was back-lit with an LED light source. The resulting spatial resolution was 0.71$mm$ X 0.71$mm$. Glass particles were tracked using an in-house code, PTVResearch \cite{Fullmer2020} based on optical flow equations. Optical distortions were removed using calibrated grid and dewarping \cite{Higham2019}, and outliers were detected by proper orthogonal decomposition \cite{Higham2016}. {\av In a recent effort \cite{Weber2021}, our method was cross-validated with other particle tracking algorithms when applied to a fluidized bed system, and the predicted velocities compared well. We remark on a few noteworthy limitations of our apparatus. Particle tracks are lost when they enter bubbles, where they become out-of-focus due to back-lighting. Only particles tracked during the entire duration of our experiments are considered for statistics to avoid unintended bias. Also, the unit is prone to slugging from tight confinement along its depth, which prohibits exploring higher pulsing frequencies and amplitudes as well as using unperturbed flow at the inlet. Further details regarding the experiments can be found in \citet{Higham2020}.}


\begin{figure}[!htpb]
\begin{center}
\includegraphics[width=0.8\linewidth]{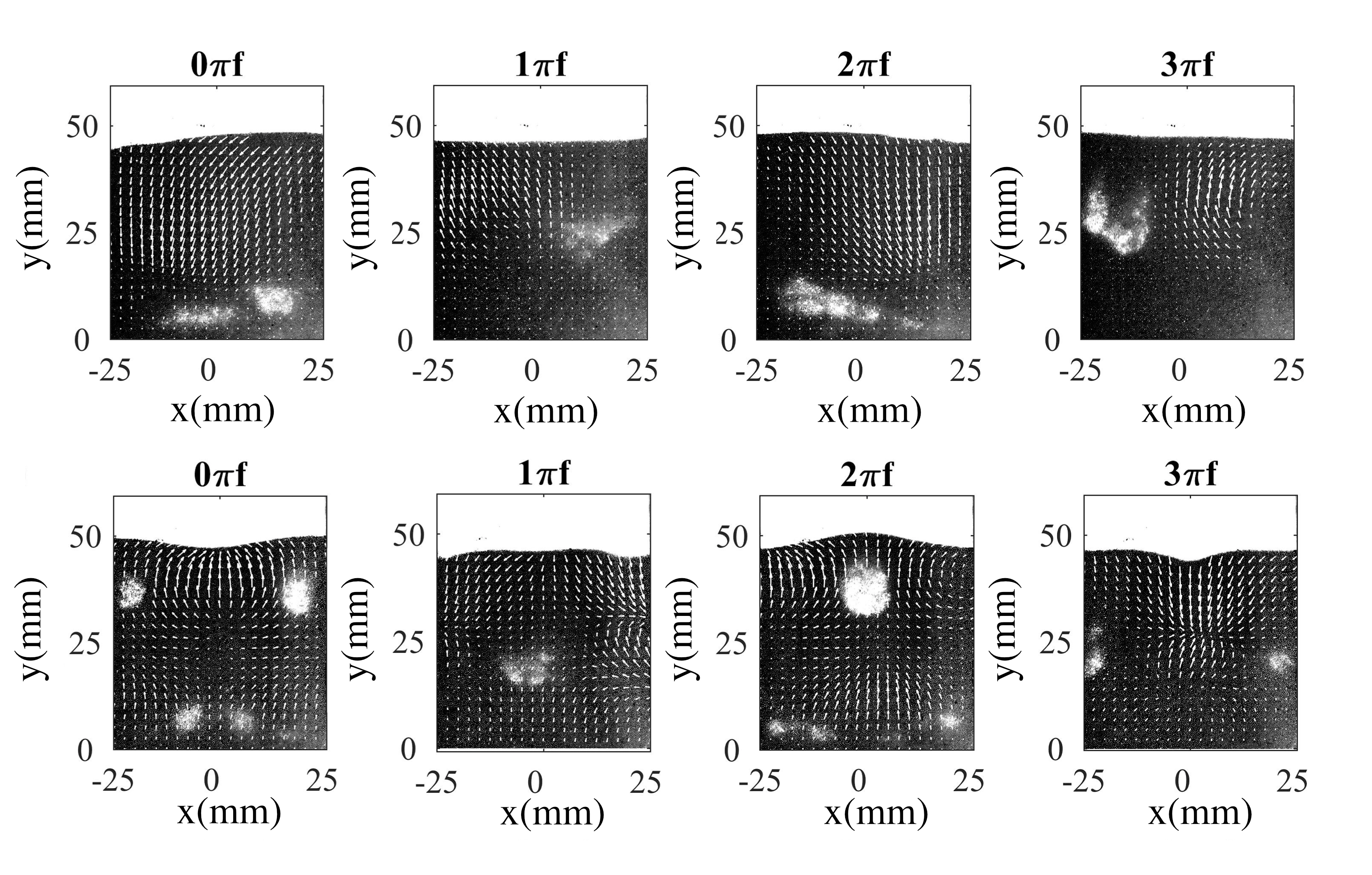}
\end{center}
\caption{Structured bubble patterns at f=4$Hz$ (top) and 6$Hz$ (bottom).}
\label{fig:Schematic2}
\end{figure}

{\av Meso-scale responses to pulsing conditions are shown in Figure ~\ref{fig:Schematic2}. Kinematic waves originate as one-dimensional planar disturbances and transition into structured bubbles as a consequence of interactions between harmonic and sub-harmonic modes. The recurring patterns are sustained provided their wavelength, $\lambda$ fit the lateral dimension. Bubbles shift by $\lambda/2$ between successive cycles. We notice $\lambda$, associated with bubble size, reduces while changing f from 4$Hz$ to 6$Hz$. At 4$Hz$, bubbles are larger and switch sides every half cycle. Wakes experience a greater compressive stress when defluidized, which makes bubbles less stable and deformed as they propagate upward. The pattern changes to a bubble at the center and two simultaneous bubbles along the walls at 6$Hz$, having a smaller size and a more distinct interface.}

\section{\label{sec:level3}Results}
\begin{figure}[!htpb]
\begin{center}
\begin{subfigure}[b]{0.4\linewidth}
    \includegraphics[width=\linewidth]{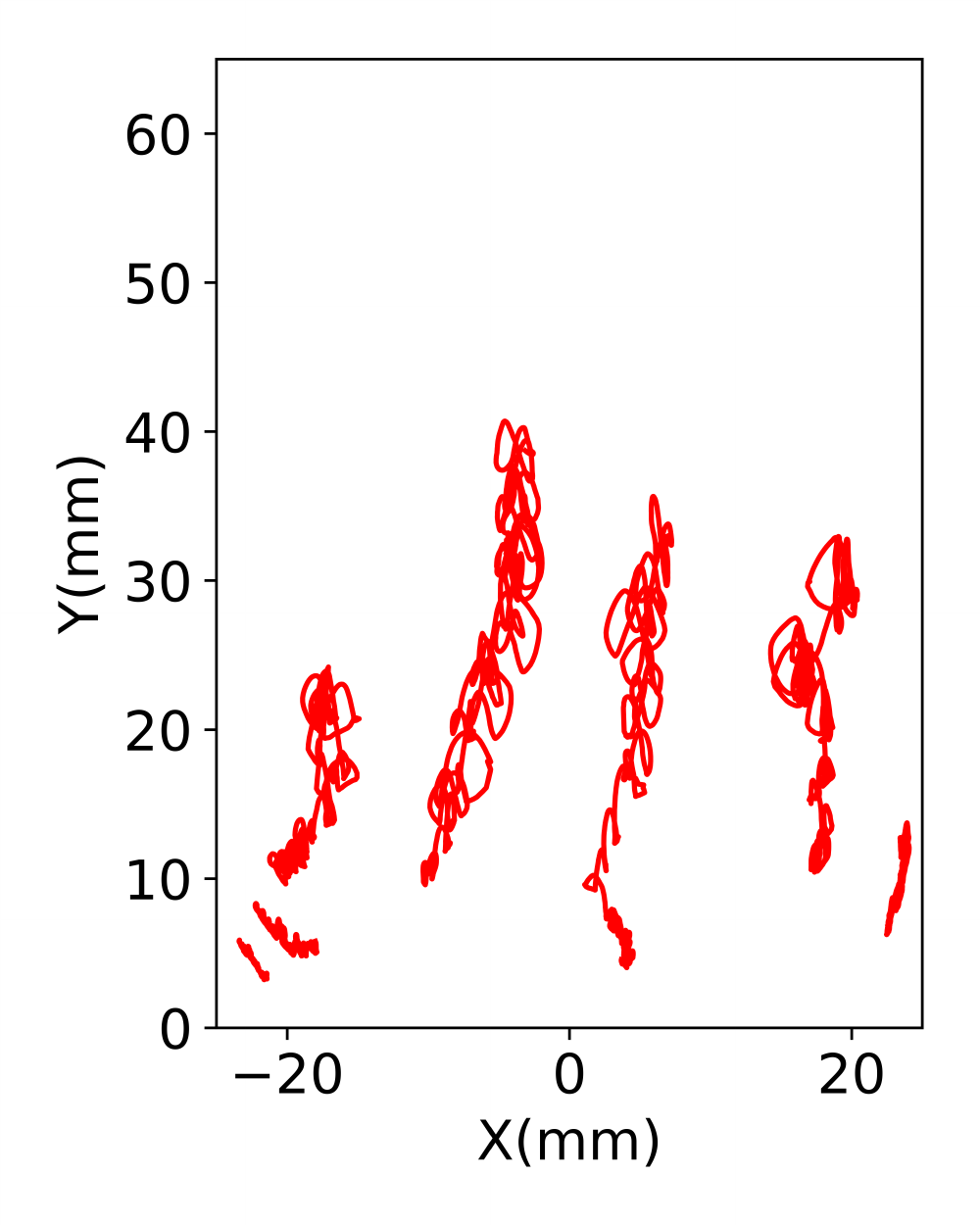}
\end{subfigure}
\begin{subfigure}[b]{0.4\linewidth}
    \includegraphics[width=\linewidth]{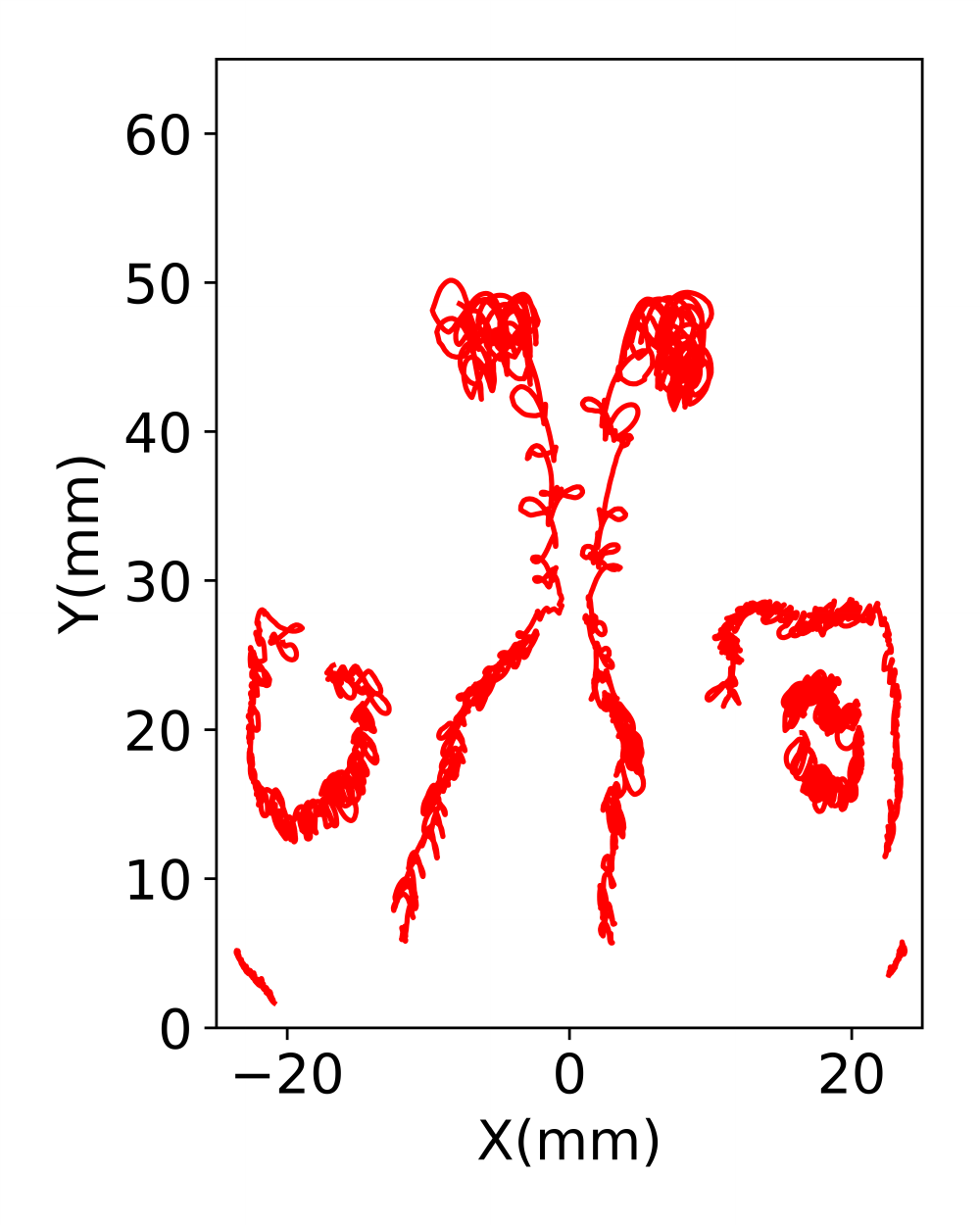}
\end{subfigure}
\end{center}
\caption{Sample trajectories at f=4$Hz$ (left) and 6$Hz$ (right).}
\label{fig:trajectory}
\end{figure}

Sample trajectories (Figure ~\ref{fig:trajectory}) indicate non-uniform diffusion in the PFB. We notice a few particles transported over much shorter distances during the entire measurement period. Even if a single particle track is considered, the motion is altered significantly depending on instantaneous location. Particles in the wake of bubbles take longer steps, while they undergo much shorter displacements in quasi-static regions. Their motion is confined by walls in the lateral direction, and the balance between gravity and inter-phase drag governs their streamwise transport. The dynamics are strongly coupled to the fluidizing medium, a mechanism neither trivial nor explicitly modeled while describing anomalous diffusion. It is worth mentioning that the effect of drag was included in the generalized Langevin equation \cite{garzo2012} to derive kinetic theory, more appropriate for fluidized granular media. Albeit, continuum modeling efforts have failed to reproduce structured bubbles in PFBs \cite{de2018universal}. Stress field in granular medium is not adequately represented by the existing frictional models, which is critical to sustain the recurring pattern. {\av We also notice spiral trajectories possibly due to strongly correlated directional changes \cite{sadjadi2015persistent}. In our case, this is caused by a combination of unidirectional forcing, lateral confines and preferential movement of bubbles. Momentum transfer from particle interactions is subdued compared to anisotropic wake-induced transport which breaks the symmetry in turning-angle and forms clockwise or counterclockwise patterns close to walls. However, diffusion characteristics vary spatially and spiraling motion is not present throughout the domain.}


\begin{figure}[!htpb]
\begin{center}
\begin{subfigure}{0.7\linewidth}
    \includegraphics[width=\linewidth]{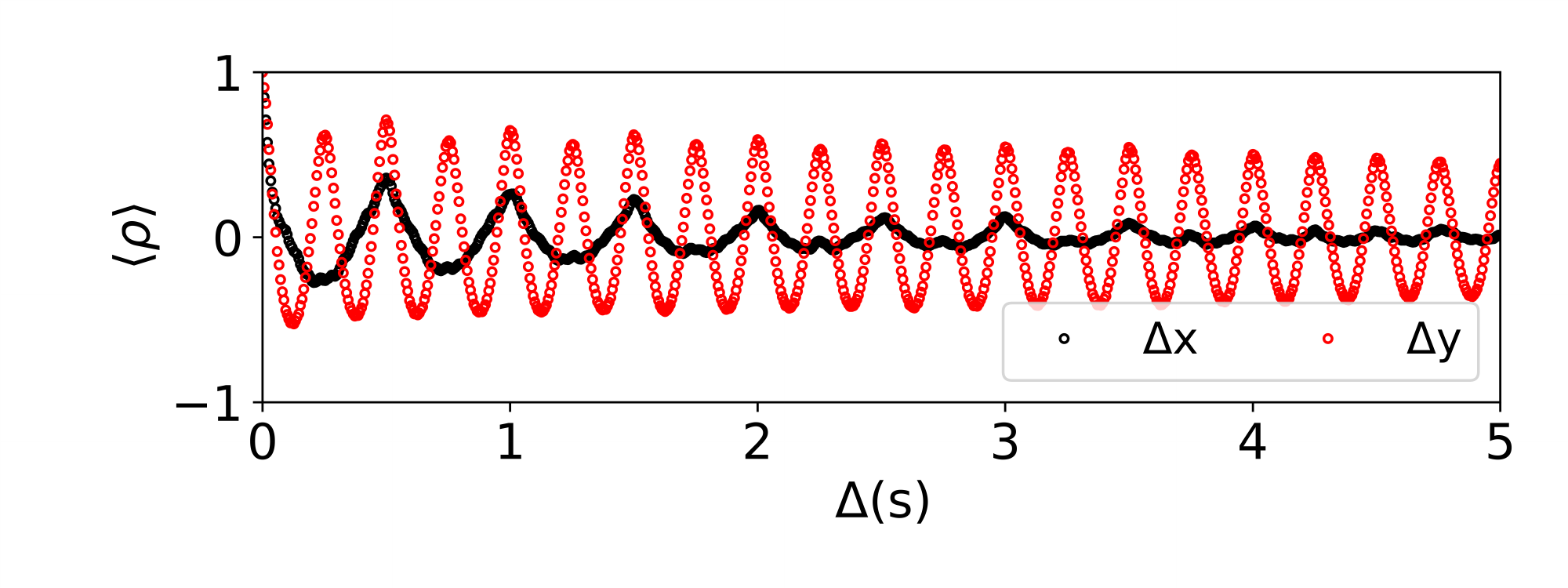}
\end{subfigure}
\begin{subfigure}{0.7\linewidth}
    \includegraphics[width=\linewidth]{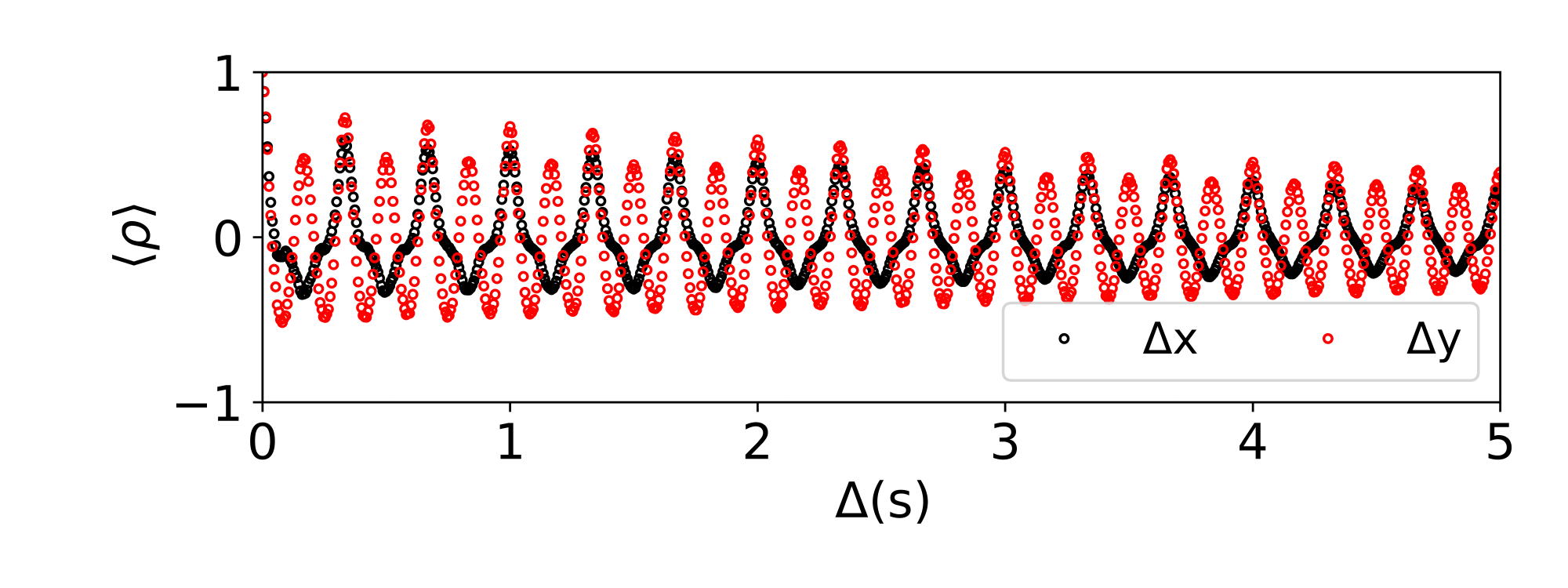}
\end{subfigure}
\end{center}
\caption{Autocorrelation of displacements, $\Delta \mathrm{x}$ and $\Delta \mathrm{y}$ at f=4$Hz$ (top) and 6$Hz$ (bottom).}
\label{fig:ACF}
\end{figure}

Next, we look at autocorrelation, $\mathrm{\rho}$ between displacements in the cartesian directions. We use the following definition,

\begin{equation}
\label{eq:ACF}
    \rho (\Delta) = \frac{\mathbf{E}\big[\Delta \mathrm{x}_{j}(t) \Delta \mathrm{x}_{j}{(t+\Delta)\big]}}{\sqrt{\mathbf{Var}\big[\Delta \mathrm{x}_{j}(t)\big] \mathbf{Var}\big[\Delta \mathrm{x}_{j}(t+\Delta)\big]}}
\end{equation}

\noindent where, $\Delta$ and $\Delta \mathrm{x}_{j}$ are lag time and particle displacement. This is ensemble-averaged to obtain $\mathrm{\langle \rho \rangle}$ shown in Figure ~\ref{fig:ACF}. Again, we notice dominant harmonic and sub-harmonic responses at both the pulsing frequencies. Lateral steps show a rapid decay of $\mathrm{\langle \rho \rangle}$ at f=4$Hz$, while the two components reveal comparable persistent memory at f=6$Hz$. This occurs in conjunction with redistribution of energy between harmonic and sub-harmonic modes as explained using proper orthogonal decomposition in our previous study \cite{Higham2020}. The long-range correlation observed in the collective behavior of particles follows the idea of Kac \cite{Kac1956}, wherein determinism evolves in multi-particle systems governed by individual stochastic differential equations.


\begin{figure}[!htpb]
\begin{center}
\begin{subfigure}{0.5\linewidth}
    \includegraphics[width=\linewidth]{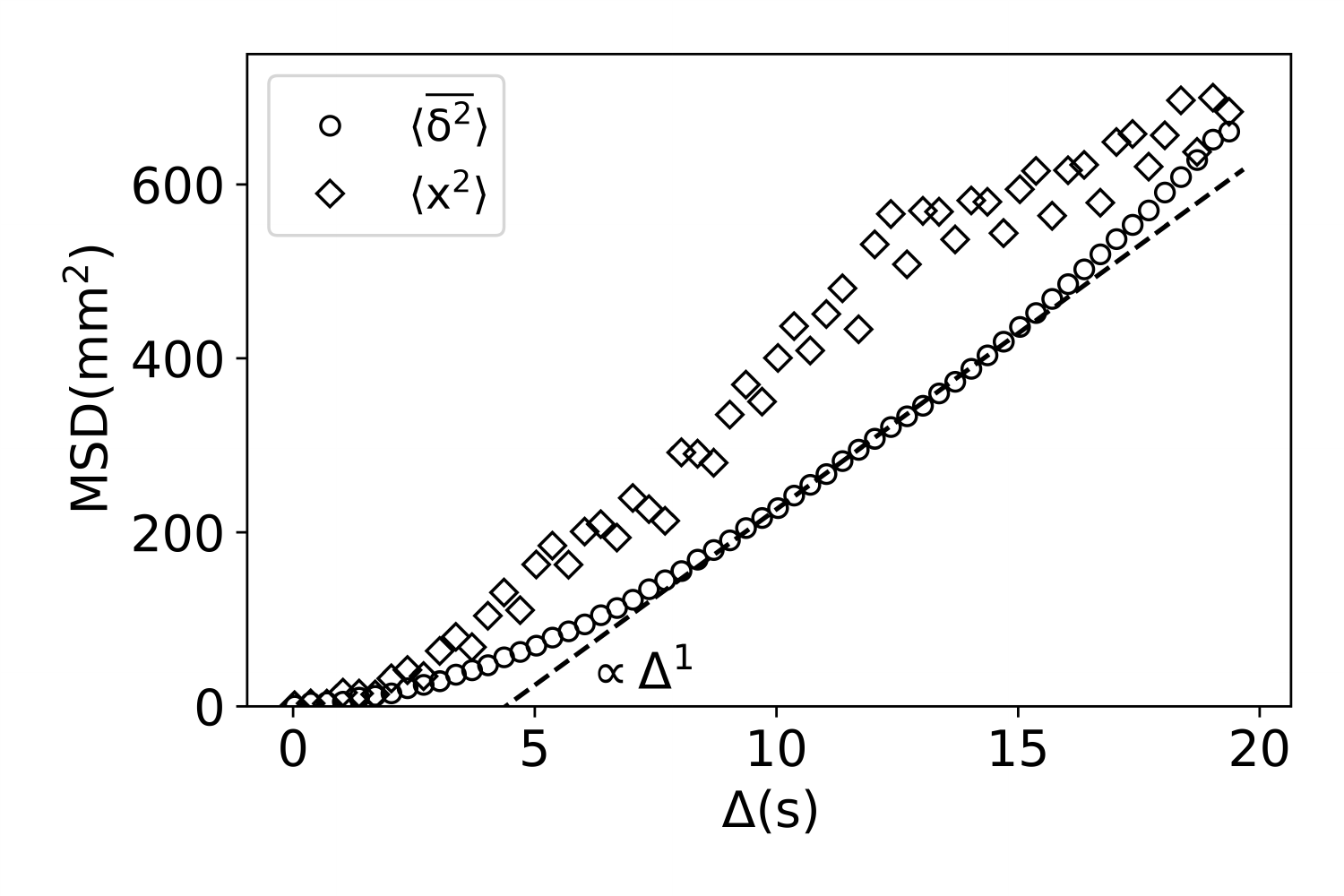}
\end{subfigure}
\begin{subfigure}{0.5\linewidth}
    \includegraphics[width=\linewidth]{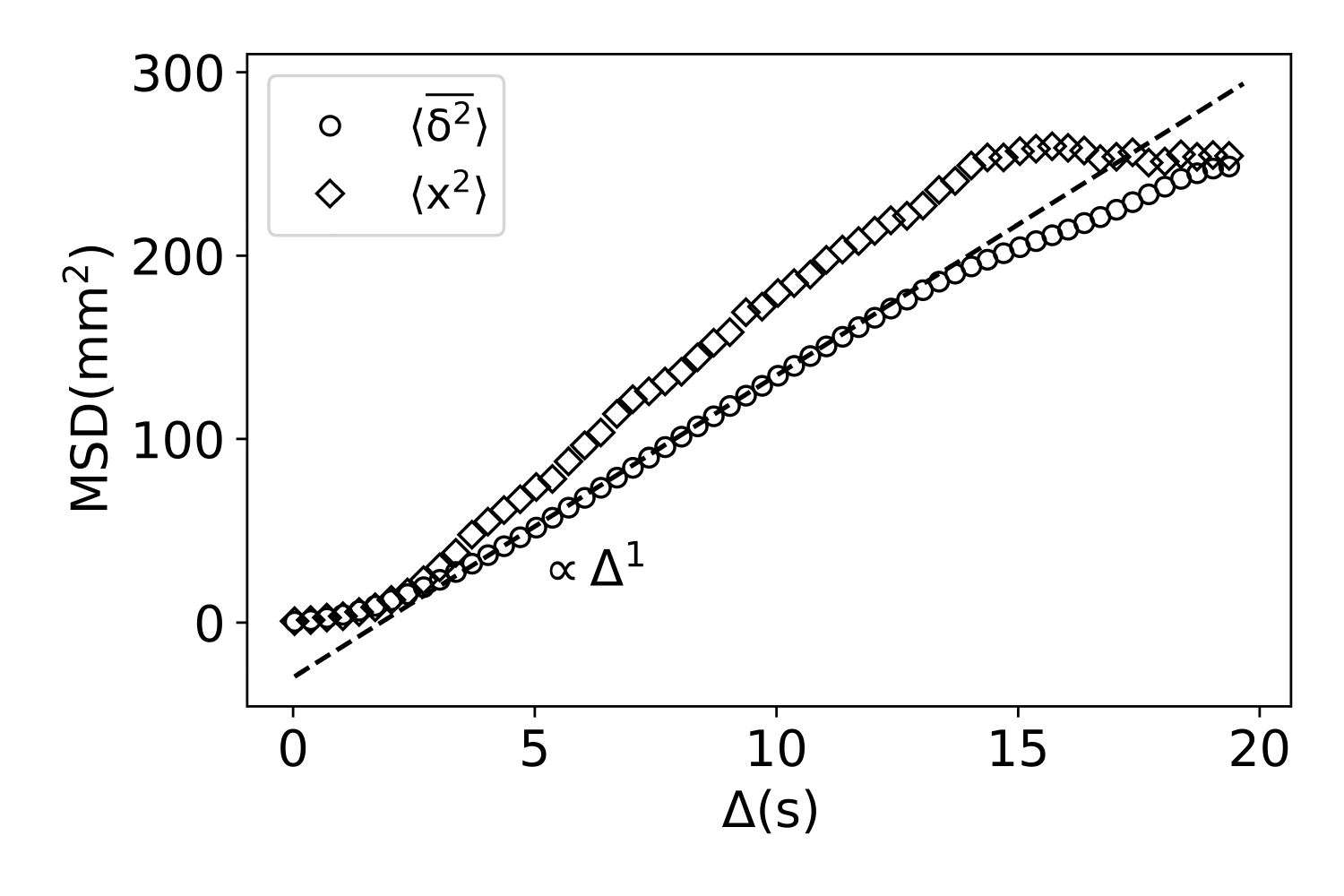}
\end{subfigure}
\end{center}
\caption{MSD from experiments at f=4$Hz$ (top) and 6$Hz$ (bottom). Dashed lines represent $\Delta^1$ sacling.}
\label{fig:MSD}
\end{figure}

~\noindent We then examine the behavior of MSD, typically used to study diffusion processes. At this point, ergodicity is not known, and we use two different measures of MSD. First, is the ensemble-averaged MSD defined as,

\begin{equation}
\label{eq:MSD1}
    \langle \mathrm{x^{2}}(\Delta) \rangle = \frac{1}{N} \sum_{i=1}^{N} |\mathrm{x_{i}}(\Delta) - \mathrm{x_{i}}(0)|^{2}
\end{equation}

\noindent where, N is the total number of tracked particles. The second measure is given by,

\begin{equation}
\label{eq:MSD2}
    \langle \mathrm{\overline{\delta^{2}(\Delta)}} \rangle = \frac{1}{\mathrm{T}-\Delta} \int_{0}^{\mathrm{T}-\Delta} \langle |\mathrm{x}(t+\Delta) - \mathrm{x}(t)|^{2} \rangle \mathrm{d}t
\end{equation}

\noindent which involves both time-averaging and ensemble-averaging, and T is the total duration of experiments. {\av $\langle \mathrm{x^{2}} \rangle$ has a nonunique scaling exponent (Figure ~\ref{fig:MSD}). Dynamically ordered bubbles result in spatially varying diffusion which may not be apparent while probing the ensemble behavior. $\langle \mathrm{\overline{\delta^{2}}} \rangle$ has a sublinear exponent initially followed by a cross over to a linear trend at long time scales similar to diffusing insulin granules \cite{tabei2013intracellular}. More tracked particles participate in wake transport over longer periods. Propagation of bubbles separates such entrained particles from crowded surroundings and their displacements become increasingly uncorrelated in time. This could cause an anti-persistent motion leading to sub-diffusion at short time scales with a gradual transition to memoryless diffusion typical of a CTRW. The final cross over to $\sim \Delta^{0}$ for $\Delta \to T$ is due to confinement as reported for GDPs and SBMs \cite{Metzler2014,Metzler2015} in contrast to purely sub-diffusive CTRWs, where plateaus are not present for time-averaged MSDs. Detailed measurements such as turning-angle distributions \cite{shaebani2014anomalous,sadjadi2015persistent,shaebani2019transient} may be required to formulate a model describing these trends in $\langle \mathrm{x^{2}}$ and $\langle \mathrm{\overline{\delta^{2}}} \rangle$.}


Besides, we notice significant difference in scaling between $\langle \mathrm{x^{2}} \rangle$ and $\langle \mathrm{\overline{\delta^{2}}} \rangle$ indicating weak non-ergodicity \cite{Bel2005,Metzler2007,Sokolov2012,Metzler2014}. This eliminates the possibility of ensemble diffusion in PFB governed by transport on a fractal support, which is ergodic by definition. $\langle \mathrm{x^{2}} \rangle$ and $\langle \mathrm{\overline{\delta^{2}}} \rangle$ show the same limiting behavior at $\Delta/T \to 0$ and $\Delta/T \to 1$. The latter is apparent from Equation ~\ref{eq:MSD2}, which has a singularity for $\Delta \to T$, thus placing the constraint, $\langle \mathrm{x^{2}} \rangle = \langle \mathrm{\overline{\delta^{2}}} \rangle$. Also, the linear scaling in $\langle \mathrm{\overline{\delta^{2}}} \rangle$ is prevalent for an appreciable period at both the values of f, previously observed for sub- and super-difussive unconfined HDPs \cite{Metzler2014}. This might lead to a false impression of Brownian motion unless supported by complementary statistical measures.


To elucidate non-ergodic dynamics in PFB, we define ergodicity-breaking parameter (EB) as,

\begin{equation}
\label{eq:EB}
\mathrm{EB(\Delta)} = \frac{\langle \overline{\delta^{2}(\Delta)}^2 \rangle}{\langle \mathrm{\overline{\delta^{2}(\Delta)}} \rangle^2} - 1
\end{equation}

\begin{figure}[!htpb]
\begin{center}
\begin{subfigure}{0.5\linewidth}
    \includegraphics[width=\linewidth]{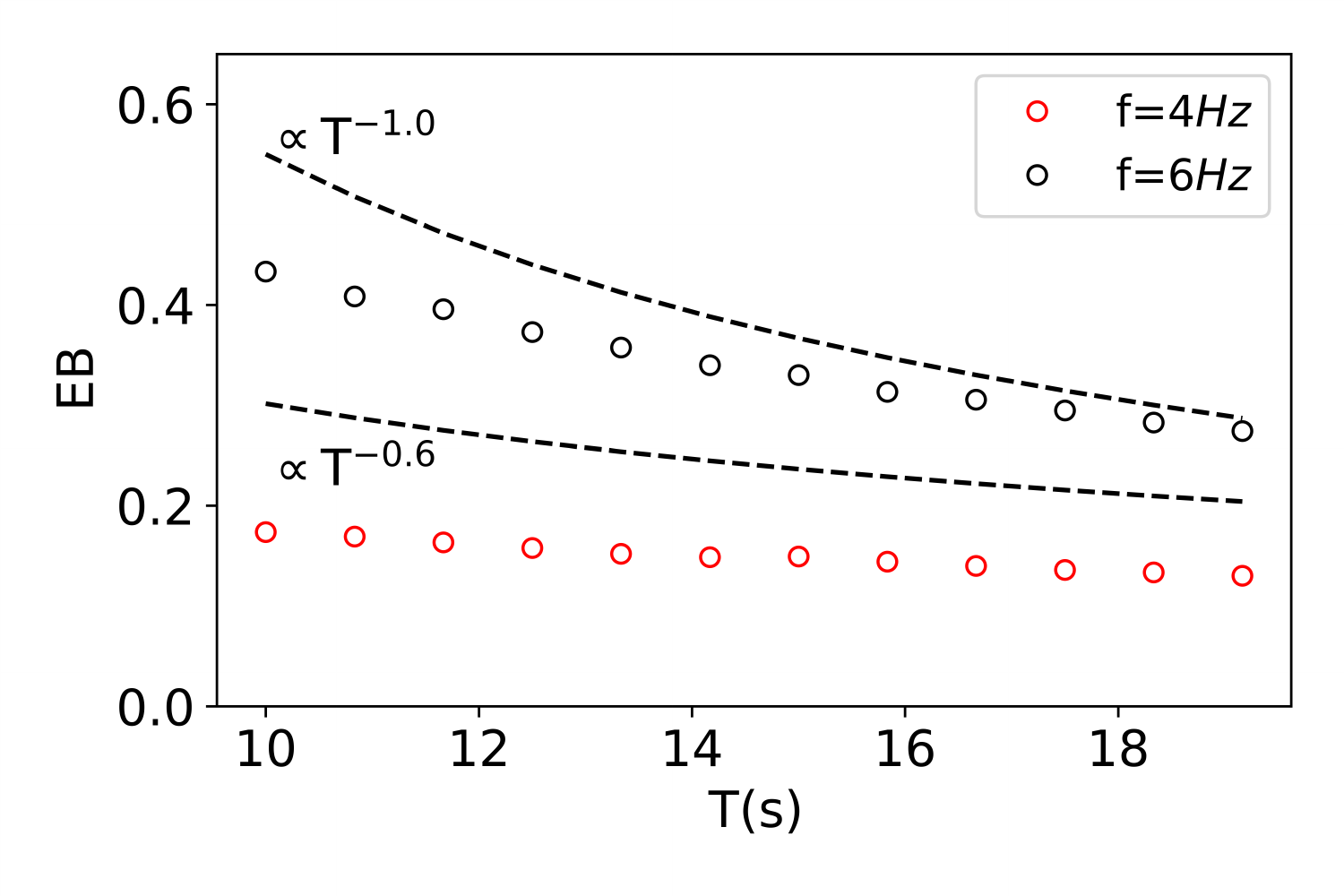}
\end{subfigure}
\begin{subfigure}{0.5\linewidth}
    \includegraphics[width=\linewidth]{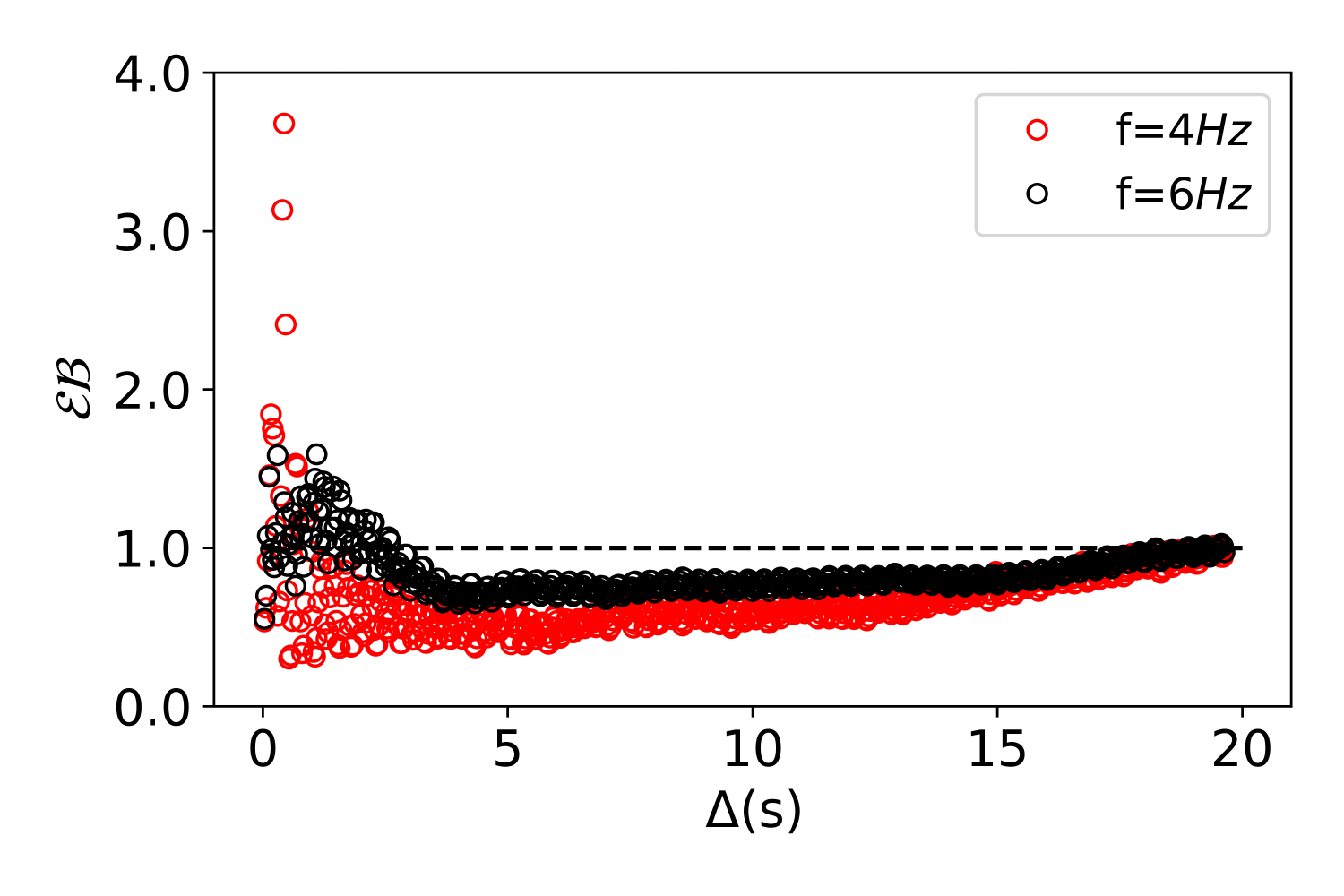}
\end{subfigure}
\end{center}
\caption{Top panel shows EB at f=4$Hz$ and 6$Hz$ using $\Delta$=0.017$s$. Dashed lines represent different slopes for guidance. The bottom panel shows variation in $\mathcal{EB}$.}
\label{fig:EB}
\end{figure}

\noindent $\mathrm{EB}$ represents dispersion in $\mathrm{\overline{\delta^{2}}}$, and we examine its variation with T to identify deviation from ergodic behavior. $\mathrm{EB}$ for a Brownian motion follows $\lim_{\Delta/T \to 0} \mathrm{EB_{BM}(\Delta)} = \frac{4}{3}\frac{\Delta}{T}$, indicated by the curve $\propto \mathrm{T}^{-1}$ in Figure ~\ref{fig:EB}. We observe a more gradual change in EB approaching a finite value for $\Delta/T \to 0$ as reported for anomalous stochastic processes governed by HDPs and CTRWs \cite{Metzler2015ChemPhys}. To further investigate the nature of ergodicity breaking, we use alternative ergodic parameter, $\mathcal{EB}$ following the definition of \citet{Metzler2013} given by,

\begin{equation}
\label{eq:EB2}
\mathrm{\mathcal{EB}(\Delta)} = \frac{ \langle \mathrm{\overline{\delta^{2}(\Delta)}} \rangle}{\langle \mathrm{x^{2}}(\Delta) \rangle}
\end{equation}

\noindent At short time scales, there is a pronounced scatter in $\mathcal{EB}$. Particles reside in a given state for a duration determined by the spatio-temporal evolution of the system. As measurement time increases, more particles transition between states and the change in $\mathcal{EB}$ becomes more moderate. We notice $\mathcal{EB} \neq 1$ at intermediate time scales, confirming deviation from ergodic dynamics. We also find $\mathcal{EB}=1$ for $\Delta/T \to 1$ due to confinement, which is not indicative of ergodicity though it mathematically represents a necessary condition. A sufficient condition for ergodicity is $\mathrm{EB} \to 0$ for $\Delta/T \to 0$ which is clearly not satisfied here.

\begin{figure}[!htpb]
\begin{center}
\begin{subfigure}{0.5\linewidth}
    \includegraphics[width=\linewidth]{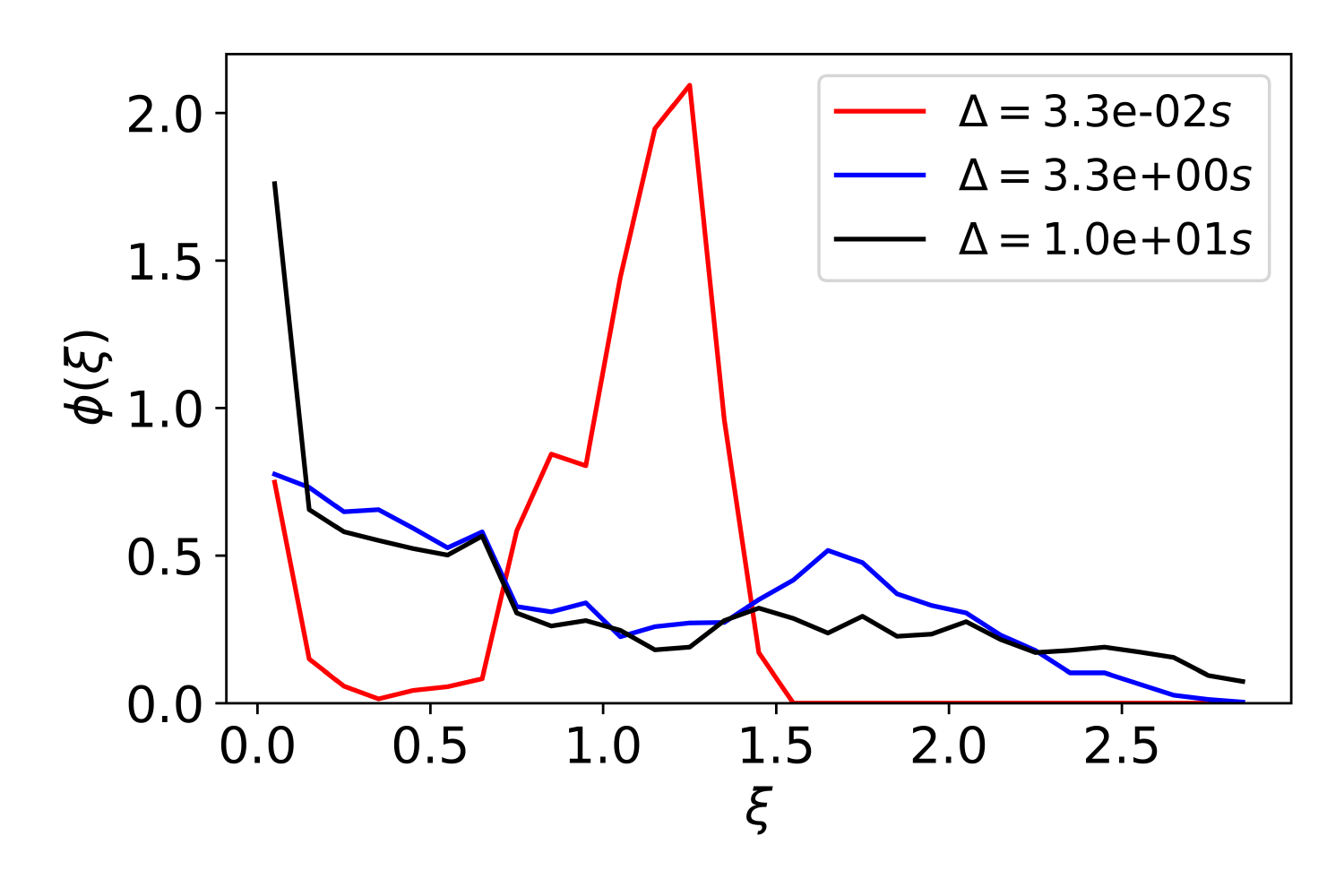}
\end{subfigure}
\begin{subfigure}{0.5\linewidth}
    \includegraphics[width=\linewidth]{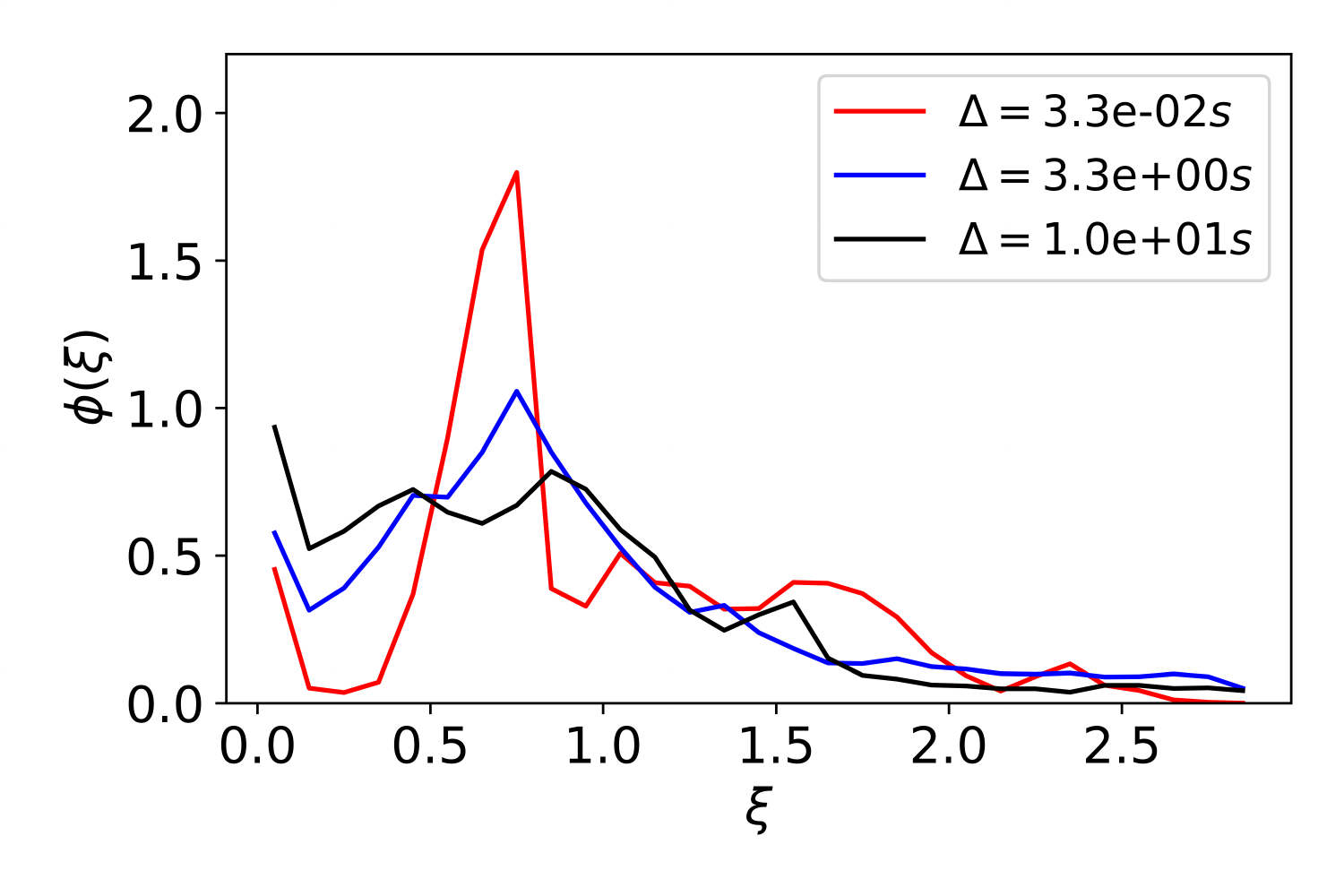}
\end{subfigure}
\end{center}
\caption{Histograms of $\xi$ from experiments at f=4$Hz$ (top) and 6$Hz$ (bottom).}
\label{fig:xi}
\end{figure}

In addition, we examine the spread in $\overline{\delta^{2}}$ (Figure ~\ref{fig:xi}) using the non-dimensional parameter $\xi$ defined as,

\begin{equation}
    \xi = \frac{\overline{\delta^{2}(\Delta)}}{\langle \mathrm{\overline{\delta^{2}(\Delta)}} \rangle}
\end{equation}

\noindent The distribution, $\phi(\xi)$ is Gaussian centered at 1 ($\xi$ = 1 represents ergodicity) for a Brownian walker. For $\Delta/T \ll 1$, $\phi(\xi)$ has distinct peaks at $\xi > 1$ (f=4$Hz$) and $\xi < 1$ (f=6$Hz$). We notice a significant scatter in $\xi$ for growing lag times while deviating from ergodic dynamics. $\phi(\xi)$ is finite at $\xi$=0 for all values of $\Delta$, as observed for sub-diffusive CTRWs \cite{Cherstvy2013}. The contribution from quasi-static particles is more prominent for longer lag times at f=4$Hz$. These findings corroborate  combination of parent stochastic processes dictating the underlying anomalous diffusion. {\av Weak ergodicity breaking essentially results from information content in single-particle trajectories which are not retained while ensemble-averaging. Besides, wake transport is not truly reproducible and distances over which it occurs could vary leading to non-equilibrium relaxation.}

\begin{figure}[!htpb]
\begin{center}
\begin{subfigure}{0.5\linewidth}
    \includegraphics[width=\linewidth]{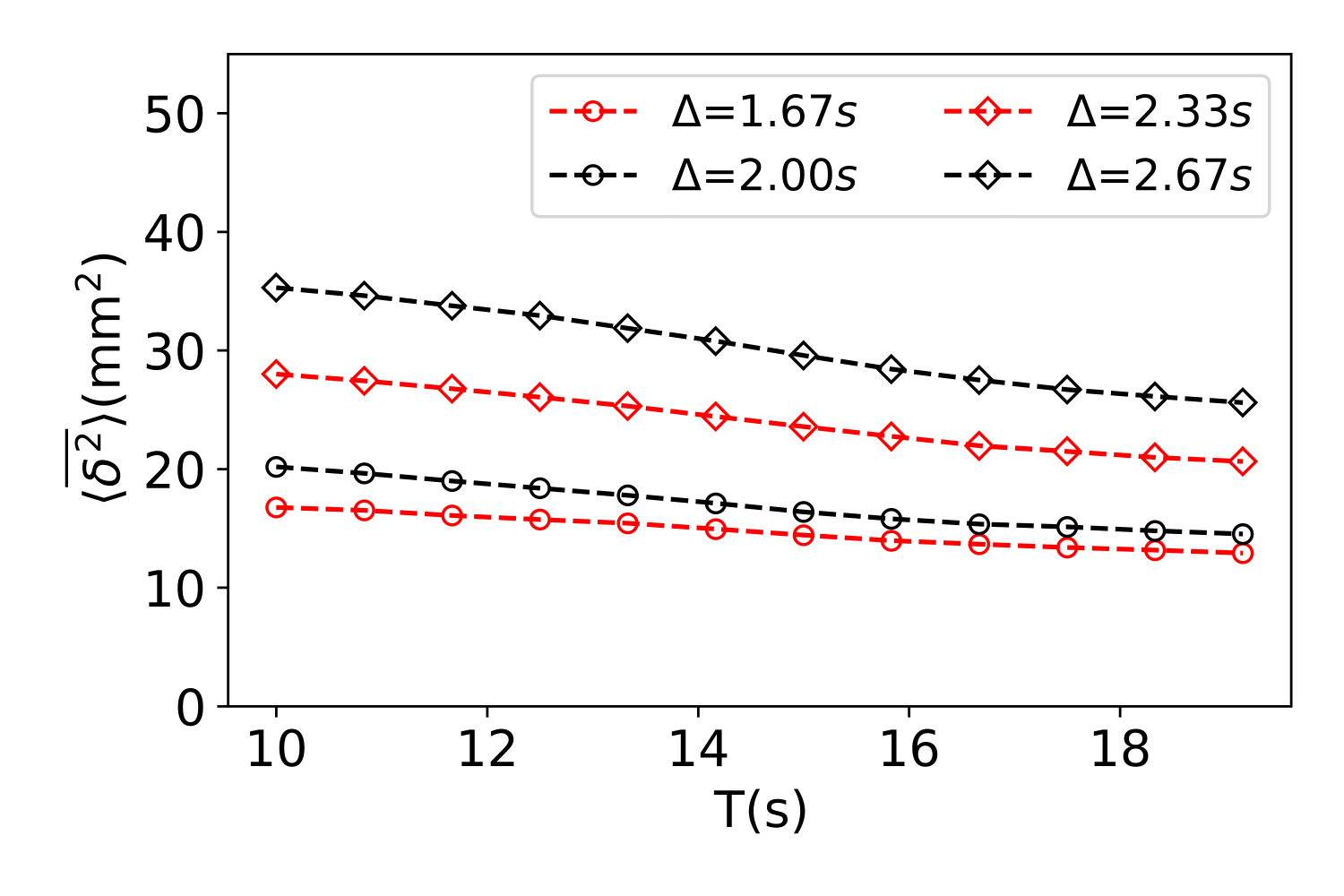}
\end{subfigure}
\begin{subfigure}{0.5\linewidth}
    \includegraphics[width=\linewidth]{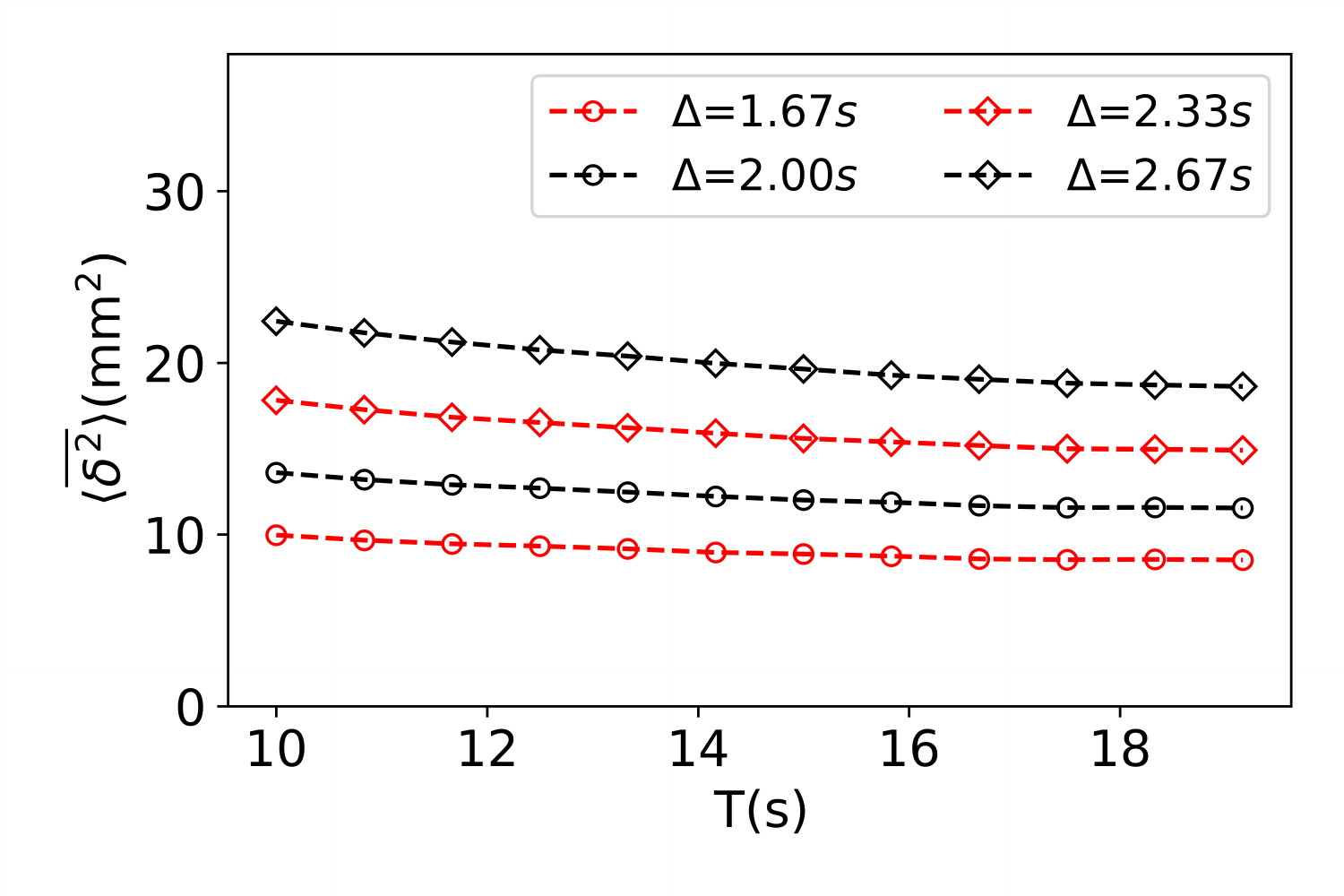}
\end{subfigure}
\end{center}
\caption{Variation of $\langle \overline{\delta^{2} \rangle}$ with measurement time at f=4$Hz$ (top) and 6$Hz$ (bottom).}
\label{fig:ageing}
\end{figure}

\begin{figure}[!htpb]
\begin{center}
\begin{subfigure}{0.5\linewidth}
    \includegraphics[width=\linewidth]{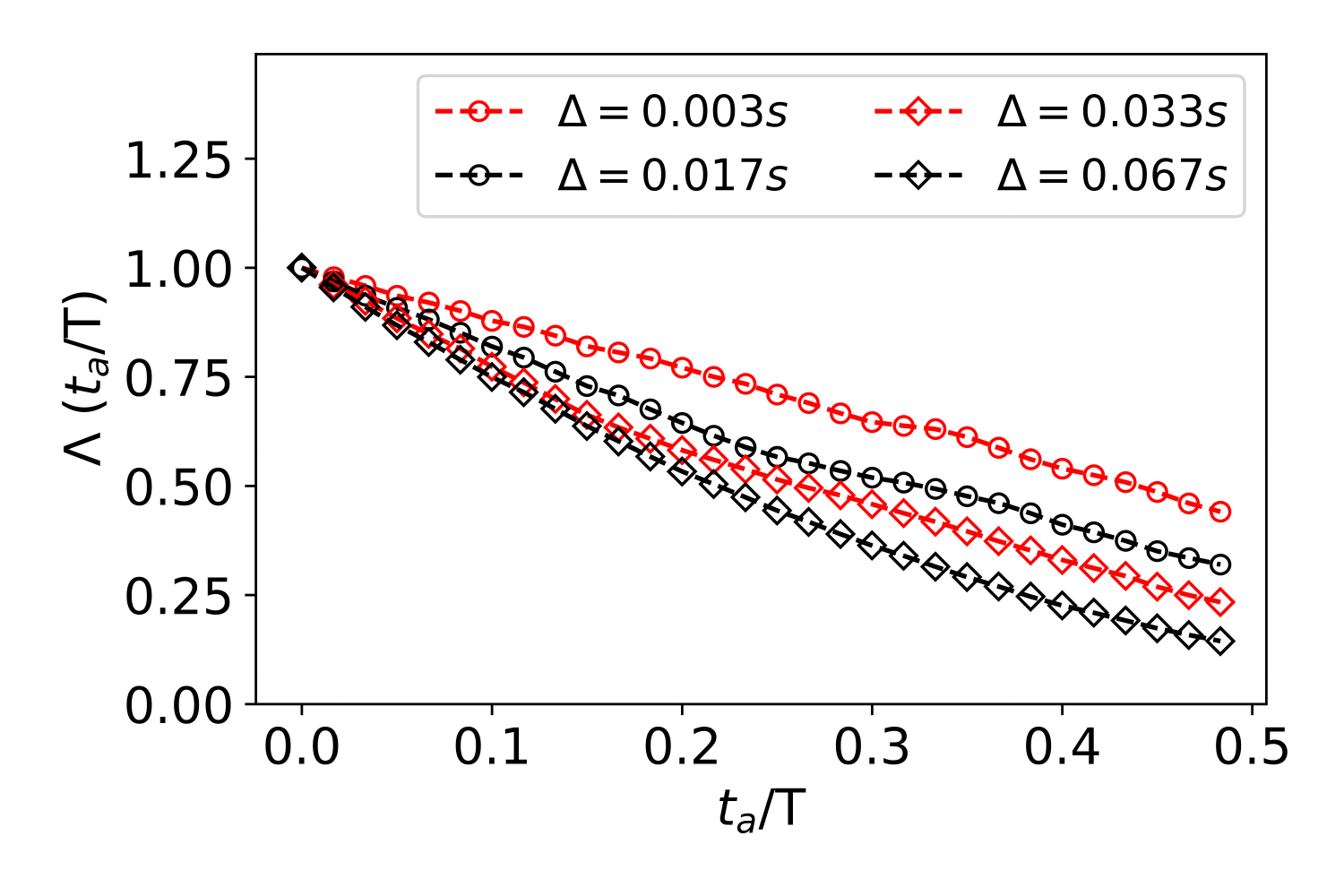}
\end{subfigure}
\begin{subfigure}{0.5\linewidth}
    \includegraphics[width=\linewidth]{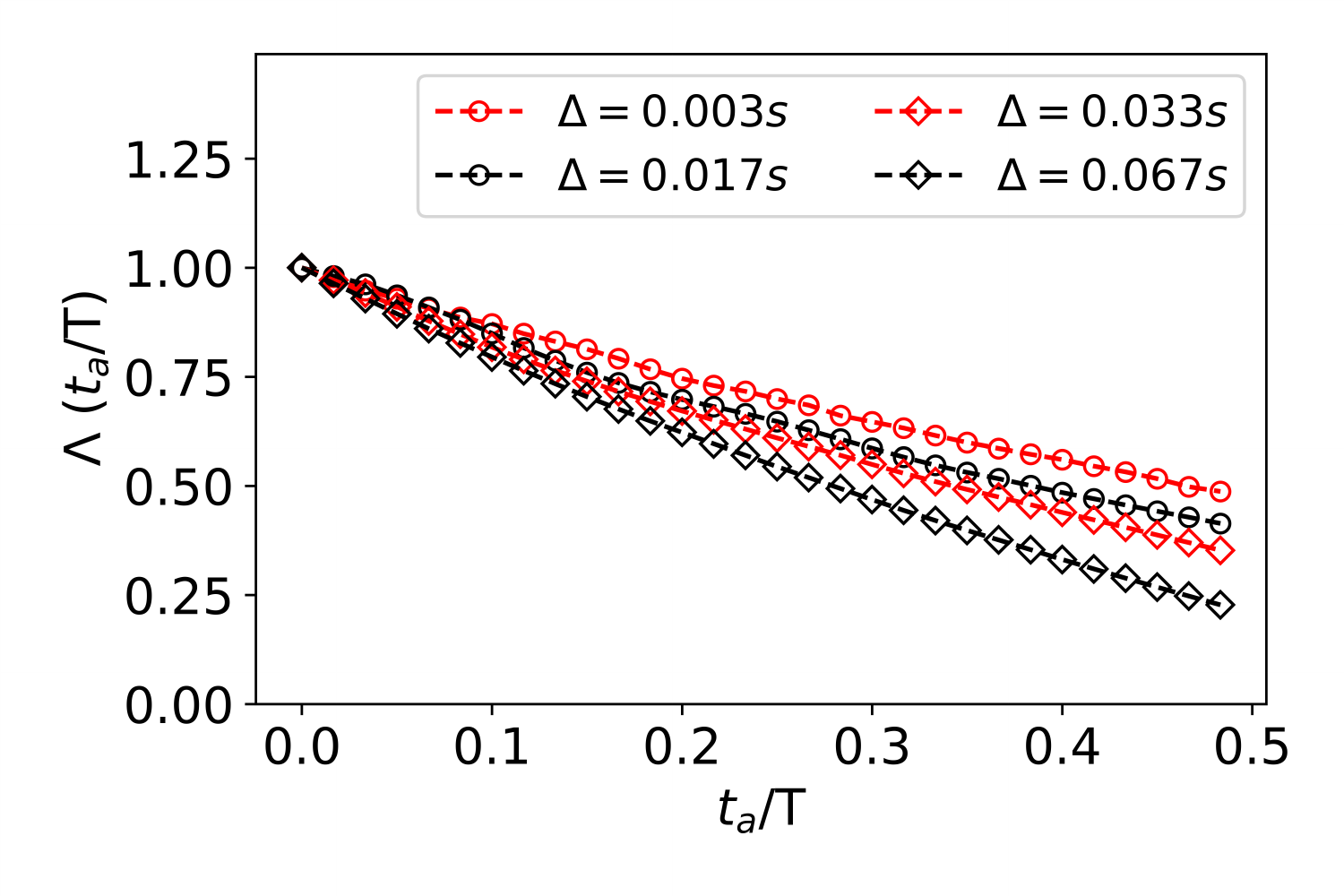}
\end{subfigure}
\end{center}
\caption{Variation of $\Lambda$ with dimensionless ageing time at f=4$Hz$ (top) and 6$Hz$ (bottom) for different values of $\Delta$ as indicated.}
\label{fig:ageingfactor}
\end{figure}

{\av Finally we look at ageing characteristics, which along with ergodicity breaking determine the (non-)stationary nature of a stochastic process. $\langle \mathrm{\overline{\delta^{2}}} \rangle$ shows a monotonic drop as a function of T (Figure ~\ref{fig:ageing}) for different values of $\Delta$. This is indicative of a collective sub-diffusive anomalous process. Analogous ageing behavior in a sub-diffusive environment is found in other instances including plasma membranes \cite{Krapf2011} and fibrin matrices \cite{aure2019damped}. As a consequence, the system appears less diffusive as it evolves longer. More and more particles transition through quasi-static regions which overpopulate the tails of wait-time distributions. If the on-off response of particles switching between wake transport and interaction with neighbors (through friction and collision) is considered, the occurrence and duration of these events appear random at short time scales. Upon prolonged measurements, occasionally long on and off states are obtained, characteristic of non-stationary and out-of-equilibrium dynamics \cite{bertin2003linear,burov2007occupation}. This in essence results in the observed ageing behavior. But individual motion of particles could vary depending on localized states. We further quantify the ensemble behavior using ageing factor \cite{Metzler2014} defined as,

\begin{equation}
\label{eq:ageingfactor}
    \Lambda(t_{a},\Delta) = \frac{ \langle \mathrm{\overline{\delta_{a}^{2}(\Delta)}} \rangle}{\langle \mathrm{\overline{\delta^{2}(\Delta)}} \rangle}
\end{equation}

\noindent $\overline{\delta_{a}^{2}}$ refers to the time-averaged MSD considering the ageing time, $t_{a}$ given by,

\begin{equation}
\label{eq:agedmsd}
    {\av \overline{\delta_{a}^{2}(t_{a})} =  \frac{1}{\mathrm{T}-\Delta-t_{a}} \int_{t_{a}}^{\mathrm{T}-\Delta} \langle |\mathrm{x}_{i}(t+\Delta) - \mathrm{x}_{i}(t)|^{2} \rangle \mathrm{d}t}
\end{equation}

\noindent The above expression is ensemble-averaged while calculating $\Lambda$ for different values of $t_{a}$ shown in Figure ~\ref{fig:ageingfactor}. There is a steady drop in $\Lambda$ even for $\Delta/T \ll 1$ due to continued localization of particles in quasi-static regions, again indicative of sub-diffusive behavior. Even though the meso-scale response (bubble pattern) is completely different at f=4$Hz$ and 6$Hz$, similar anomalous diffusion characteristics are observed.}


\section{\label{sec:level4}Conclusion}

We analyzed anomalous diffusion in PFB having a spatio-temporal dependence using single-particle tracking, and found traits from a combination of parent stochastic processes. PFB represents a driven granular system having complex non-linear interactions. Finite memory or long-range correlations stem from individual stochastic motions, in line with the ideas of Kac surrounding propagation of chaos \cite{Kac1956}. Time-averaged and ensemble-averaged MSDs deviate indicating weak ergodicity breaking. MSDs approach a plateau similar to constrained GDPs or SBMs, although our setup is weakly confined wherein the streamwise transport is balanced between inter-phase drag and gravity. The distribution of amplitude scatter is wide, non-Gaussian, asymmetric, and has a finite contribution at zero stemming from particles in quasi-static surroundings, a feature prevalent in CTRWs. The system also exhibits ageing as more traps are encountered over a prolonged duration. The ageing factor decays monotonically suggesting an overall subdiffusive process at the two pulsing frequencies having a different meso-scale response altogether. We expect structured flow patterns in PFBs while lowering the effective diffusivity compared to fluidized media having an unperturbed inflow, which needs to be verified.


\bibliography{Ref}

\end{document}